\documentclass[hidelinks]{IEEEtran}

\usepackage{color}
\usepackage{algorithm}
\usepackage[utf8]{inputenc}
\usepackage{amsmath}
\usepackage{mathrsfs}
\usepackage{amsfonts}
\usepackage{amsthm}
\usepackage{setspace}
\usepackage{bbm}
\usepackage{mathrsfs}
\usepackage{amssymb}
\usepackage[noend]{algpseudocode}
\usepackage{relsize}
\DeclareFontFamily{OT1}{pzc}{}
\DeclareFontShape{OT1}{pzc}{m}{it}{<-> s * [1.10] pzcmi7t}{}
\DeclareMathAlphabet{\mathpzc}{OT1}{pzc}{m}{it}
\usepackage{tikz}
\usetikzlibrary{trees}
\usepackage{subcaption}
\usepackage{caption}
\usepackage{epsfig}
\DeclareCaptionFont{mysize}{\fontsize{8}{9.6}\selectfont}

\usepackage{bm}
\usepackage{siunitx}
\usepackage{amsmath}
\usepackage{cite}
 
\makeatletter
\renewcommand{\maketag@@@}[1]{\hbox{\m@th\normalsize\normalfont#1}}%

\DeclareMathOperator*{\argmin}{argmin}
\title{Energy-Efficient D2D-Aided Fog Computing under Probabilistic Time Constraints 
\thanks{This work was supported in part by the Natural Science and Engineering Research Council of Canada under the Discovery Grant program.}
\thanks{\copyright 2021 IEEE. Personal use of this material is permitted. Permission from IEEE must be obtained for all other uses, in any current or future media, including reprinting/republishing this material for advertising or promotional purposes, creating new collective works, for resale or redistribution to servers or lists, or reuse of any copyrighted component of this work in other works.}
\thanks{Published in the \emph{Proceedings of the 2021 IEEE Global Communications Conference (Globecom)}.}
}

\author{\IEEEauthorblockN{Onur Karatalay,
		Ioannis Psaromiligkos and Benoit Champagne}\\
	\IEEEauthorblockA{Department of Electrical and Computer Engineering,
		McGill University, Montr\'{e}al, QC, Canada.\\
		Email: onur.karatalay@mail.mcgill.ca;
		ioannis.psaromiligkos@mcgill.ca;
		benoit.champagne@mcgill.ca	}}

\IEEEoverridecommandlockouts
\begin{document}

\maketitle
\begin{abstract}
Device-to-device (D2D) communication is an enabling technology for fog computing by allowing the sharing of computation resources between mobile devices. However, temperature variations in the device CPUs affect the computation resources available for task offloading, which unpredictably alters the processing time and energy consumption. In this paper, we address the problem of resource allocation with respect to task partitioning, computation resources and transmit power in a D2D-aided fog computing scenario, aiming to minimize the expected total energy consumption under probabilistic constraints on the processing time. Since the formulated problem is non-convex, we propose two sub-optimal solution methods. The first method is based on difference of convex (DC) programming, which we combine with chance-constraint programming to handle the probabilistic time limitations. Considering that DC programming is dependent on a
good initial point, we propose a second method that relies on only convex programming, which eliminates the dependence on user-defined initialization. Simulation results demonstrate that the latter method outperforms the former in terms of energy efficiency and run-time.

\end{abstract}
\vspace{-.1cm}
\section{Introduction}
Major advances in wireless networking technologies enabling ultra-high data rates with low latency have led to the proliferation of computation intensive applications, such as augmented reality, interactive gaming and video streaming \cite{YLIU}. However, considering the exacting computation and storage requirements of these applications, the rate
of technological advancements on the device side has been generally slower, so that they can hardly meet such demands. Cloud computing, in which user devices offload their computation-intensive tasks to much more powerful remote servers, can help devices reduce their computation burden \cite{PPierleoni}. However, due to network congestion, cloud computing may not be suitable for real-time applications requiring ultra-low latency and very high bandwidth \cite{KKaur}. Mobile edge computing (MEC) provides an alternative to cloud computing by offloading computation to servers at the edge of the network, thereby reducing end-to-end delays and data processing bottlenecks \cite{XHE}. However, the computation capability of MEC servers is limited, and in situations involving extensive data traffic and high user density, some mobile devices might not be able to access them, which negatively impacts the quality of service \cite{JLI}.

As a complement to cloud and edge computing, fog computing provides a decentralized framework whereby the available computation resources of nearby mobile devices are exploited for task offloading through incentive policies \cite{MZENG,SLUO,RBeraldi}. Due to the proximity of the available resources, this type of task partitioning mechanism reduces the latency and the backbone traffic in the network, and in turn, increases energy efficiency \cite{YLAN}. Device-to-device (D2D) communication, which will play an important role in fifth generation (5G) and beyond 5G (B5G) networks, is a strong candidate to enable fog computing. In \cite{YLAN}, the authors focus on  maximization of total utility in terms of energy and time consumption in a D2D-aided fog computing scenario. In \cite{SJosilo}, minimization of average task completion time is considered by using a game theoretical model. Reference \cite{LPu} considers network-assisted D2D fog computing, in which the objective is to minimize time-average energy consumption, whereas in \cite{HXing} the problem of computation latency minimization in D2D fog computing is studied by considering an optimal task assignment strategy. However, the aforementioned studies do not consider CPU throttling during task offloading, as discussed below.

In mobile devices, dynamic thermal management (DTM) schemes control the on-chip temperature by lowering the voltage and frequency of the CPU to prevent damage in the case of high temperature \cite{JMKim}. Ideally, devices allocate the highest available CPU frequency, measured in cycles per second, to perform a given task within a minimum amount of time. Due to DTM, however, significant yet unpredictable fluctuations in allocated CPU frequency do occur \cite{OSahin}. Since real-time applications require low latency and strict processing times, a random reduction in CPU frequency negatively impacts task offloading. Consequently, to optimize fog computing performance subject to this type of uncertainty, allocation of computation resources should be treated as a probabilistic optimization problem rather than a deterministic one. 

Motivated by the aforementioned challenges, we address the problem of optimal resource allocation in terms of  task  partitioning,  computation  resources  and transmit powers, in D2D-aided mobile fog computing. Specifically, we aim to minimize the expected total energy consumption under probabilistic constraints on the task processing time. Since the formulated problem is non-convex, finding the global optimum is generally intractable; therefore, we  propose  two  sub-optimal solution methods. The  first  method  leverages the difference of convex (DC) optimization framework combined with chance-constraint  programming  to  handle  the  probabilistic constraints. Nonetheless, we find that the performance of DC programming remains sensitive to the choice of the initial  point. To overcome this difficulty, we develop a second method that relies solely on convex programming. Simulation results demonstrate that the second method outperforms the first one in  terms  of  energy efficiency and  run-time, while both methods offer significant energy savings over local computation.

The paper is organized as follows. In Section II, we describe the system model and formulate the problem statement. In Section III, we present the proposed DC and convex programming methods. Simulation results are presented and discussed in Section IV. Finally, Section V concludes the paper.

\vspace{-.11cm}
\section{System Model and Problem Statement} \label{Sec:SystemModel}
\vspace{-.11cm}
\subsection{System Model}
\begin{figure}[h]
	\centering
	\includegraphics[width=60mm]{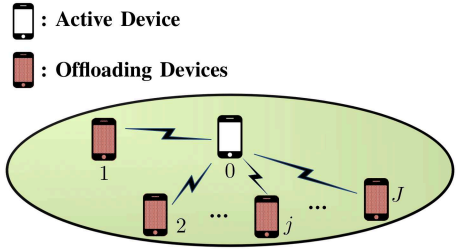}
	\small
	\caption{D2D-aided fog computing scenario, where an active device (indexed by  $0$) can offload its tasks to nearby offloading devices (indexed by $j\in \mathcal{J}$) .}
	\normalsize
	\label{Network}
\end{figure}

\vspace{-.12cm}
We consider a wireless sub-network comprised of a single active device that has a computation-intensive task to perform and $J$ offloading devices which can be used to offload this task, as seen in Fig. \ref{Network}. We label the active device by $0$ and the offloading devices by 
$j \in \mathcal{J} = \{1,2,\ldots\}$.
Similar to \cite{SJosilo}, we assume  simultaneous orthogonal side-links to establish D2D communications between the active device and each one of the offloading devices prior to task offloading.

The computation task  of the active device is characterized by the tuple $( b,c,t^{\text{max}})$. 
Here, $b$ indicates the task size in bits, $c$ denotes the number of CPU cycles required to process one bit of data, and $t^{\text{max}}$ is the maximum time limit for completing the task. 
The device may compute its task locally  and/or partially offload it to one or more offloading devices in $\mathcal{J}$. 
Accordingly, the task size can be decomposed as:
\vspace{-.02cm}
\begin{align}
 b=b_{0} + \sum_{j\in\mathcal{J}}{b_{j}} \label{DataArranged}
\end{align} 
where $b_{0}$ and $b_{j}$ represent the portions kept at the active device and sent to the $j$th offloading device, respectively. 
These portions are collected in the vector $\boldsymbol{b}=[b_{0}\ b_{1}\hspace{-0.1cm}\ \ldots \hspace{-0.1cm}\ b_{J}]^\top$ where ${}^\top$ denotes the transpose operation.

To compute the local portion of its task, the active device allocates a part $f_{0}$ of its computation resources, measured in CPU cycles per second, which cannot exceed its  maximum computation capability $ f_0^{\max}$.
However, due to unpredictable CPU throttling, e.g., resulting from temperature fluctuations, the actual computation resource used by the device is $\Tilde{f}_{0}=(1-{\xi}_0)f_{0}$, where ${\xi}_0\in[0,1]$ is a random variable with known distribution. Denoting the time it takes to complete the local portion of the task at the active device as $t^{\text{co}}_0$, we can write:
\begin{align}
\Tilde{f}_{0}t^{\text{co}}_0= b_{0}c \label{LocalCompTime}
\end{align}
The energy consumed for local computation is given by \cite{NLi}:
\begin{align}
E^{\text{loc}}=\kappa b_{0} c \Tilde{f}_{0}^2 = \kappa \Tilde{f}_{0}^3 t^{\text{co}}_0\label{LocalEnergy}
\end{align}
where $\kappa$ is an effective capacitance constant that depends on the chip architecture.

The active device also uploads to the $j$th offloading device the corresponding task portion of size $b_j$. The achievable data rate for transmission to the $j$th device is:\\*
\vspace{-.1cm}
\begin{align}
R_{j}=\hspace{.5mm} W\mbox{log}_2\bigg(1+\dfrac{ P_{j} G_{j}}{ N_0 }\bigg)
\label{ChannelCapacity}
\end{align}
where $P_{j}$ is the allocated transmission power, $G_{j}$ is the channel gain, $W$ is the channel bandwidth, and $ N_0$ is the noise power. Denoting by $t^{\text{up}}_{j}$ the upload time, we have:
\vspace{-.1cm}
\begin{align}
b_{j}=R_{j}t^{\text{up}}_{j}   \label{TaskUpload}
\end{align}
\vspace{-.55cm}

As in the case of the active device, the $j$th offloading device allocates a part $f_{j}$ of its computation resources, which cannot exceed its maximum computation capability $ f_{j}^{\max}$, to complete the offloaded task. As before, the actual computation resource used is $\Tilde{f}_{j}=(1-{\xi}_j)f_{j}$ where ${\xi}_j$ is a random variable with known distribution. 
Then, similar to (\ref{LocalCompTime}), we have:
\vspace{-.1cm}
\begin{align}
 \Tilde{f}_{j}t^{\text{co}}_{j} = b_{j}c\label{TaskCompute}
\end{align}
where $t^{\text{co}}_{j}$\hspace{-.04cm} is the time it takes the complete the offloaded portion. 

Overall, the energy consumed to upload and compute the offloaded tasks is:
\vspace{-.16cm}
\begin{align}
E^{\text{off}}&=\sum_{j\in\mathcal{J}} \big(P_{j}t^{\text{up}}_{j}+\kappa \Tilde{f}_{j}^3t^{\text{co}}_{j}\big)\label{OffloadingEnergy}
\end{align}
Finally, the total energy consumed to complete the task can be given as a sum of two terms: 
\begin{align}
E&= E^{\text{loc}} + E^{\text{off}}
=\underbrace{\sum_{j\in\mathcal{J}}\frac{P_{j}b_{j}}{R_{j}}}_{\phi(\textbf{p},\textbf{b})} + \underbrace{\sum_{i\in\mathcal{I}}\kappa b_{i} c \Tilde{f}_{i}^2}_{\psi(\textbf{b},\Tilde{\textbf{f}})}
\label{TotalEnergy}
\end{align}
where $\mathcal{I}=\{0\}\cup \mathcal{J}$, while $\phi(\textbf{p},\textbf{b})$ and $\psi(\textbf{b},\Tilde{\textbf{f}})$ are the total task uploading energy and the total task computation energy, respectively. Furthermore, $\textbf{p}=[ P_{1}\  ...\ P_{J}]^\top$ contains the transmit powers of the active device to its offloading devices and $\Tilde{\textbf{f}}=[\Tilde{f}_{0}\ \Tilde{f}_{1}\ ...\ \Tilde{f}_{J}]^\top$ contains the actual computation resources used by the devices.

\subsection{Problem Statement}
In this paper, we address the problem of optimal resource allocation, in terms of task portions, computation resources and transmit powers, in the above D2D-aided fog computing scenario. Since the allocated computation resources have a random nature, we aim to minimize the expected value of the total energy consumption subject to probabilistic constraints on the task processing times:
\vspace{-.1cm}
\begin{subequations}
\begin{align}
\mathscr{P}_1:  \min_{\textbf{p}, \textbf{b}, \textbf{f}, \textbf{t}} \quad \mathbb{E}_{\boldsymbol{\xi}}\big[E\big] \label{Prob1}\\
&\hspace{-2.72cm}\textrm{s.t.} \quad  0 \le \sum_{j\in\mathcal{J}}P_{j}\le P^\text{max}\label{C1}\\
&\hspace{-1.93cm} \sum_{i\in\mathcal{I}}{b_{i}} =b\label{C2} \\
&\hspace{-1.93cm}  \mathbb{P}(t^{\text{co}}_0\le t^{\text{max}})\ge \gamma  \label{C3} \\
&\hspace{-1.93cm}  \mathbb{P}\big( t^{\text{co}}_{j}\le\hspace{-.8mm} (t^{\text{max}}-t^{\text{up}}_{j})\big)\hspace{-.8mm}\ge \gamma , \forall j\hspace{-.8mm}\in\hspace{-.8mm}\mathcal{J} \label{C4} \\
&\hspace{-1.93cm}  b_{j}=R_{j}t^{\text{up}}_{j},\ \forall j\in\mathcal{J}\label{C5} \\
&\hspace{-1.93cm}  f_{0}\le f^{\text{max}}_0, f_{j}\le f^{\text{max}}_j, \forall j\in\mathcal{J} \label{C7} \\
\vspace*{-.1cm}
&\hspace{-1.93cm} {b}_0, b_{j}, f_{0}, f_{j}, t^{\text{up}}_{j}\ge0 \quad  \forall j\in\mathcal{J}\label{C8}
\end{align}
\end{subequations}
where $\textbf{f}=[f_{0}\ f_{1}\ ...\ f_{J}]^\top$ contains the allocated computation resources and $\textbf{t}=[t^{\text{up}}_{1}\ ...\ t^{\text{up}}_{J}]^\top$ includes the task upload times to each offloading devices. Also, $\mathbb{E}_{\boldsymbol{\xi}}[\cdot]$ is the expectation operator, and $\mathbb{P}(\cdot)$ is the probability of an event.

In problem $\mathscr{P}_1$, the expectation in (\ref{Prob1}) is taken over the distribution of the random vector $\boldsymbol{\xi} = [\xi_{0}\ \xi_{1}\ ...\ \xi_{J}]^\top$, constraint (\ref{C1}) limits the total transmit power of the active device to $P^\text{max}$ while constraint (\ref{C2}) guarantees that the task portion sizes add up to the original task size. Constraints (\ref{C3}) and (\ref{C4}) stipulate that the probability of completing the task within the maximum time limit $t^{\text{max}}$ is higher than a given reliability level $\gamma \in [0,1]$.
Constraint (\ref{C5}) ensures that the channel rate and corresponding task uploading time are consistent with the allocated task portions. Finally, constraint (\ref{C7}) indicates that the allocated computation resources cannot exceed the computation capabilities of the devices and constraint (\ref{C8}) expresses the non-negative nature of the decision variables.

\section{Proposed Sub-Optimal Methods} \label{Sec:SubOptTaskOff}
Due to the non-convex objective function (\ref{Prob1}) and the non-convex constraints (\ref{C3}), (\ref{C4}), problem $\mathscr{P}_1$ is intractable. Therefore, in this section, we propose two sub-optimal methods to solve problem $\mathscr{P}_1$.

In the first method, we write the non-convex objective function and the non-convex constraints as difference of convex functions (DCF), while using chance-constraint programming to handle the probabilistic time constraints.
The new optimization problem can then be solved using DC programming. In the second method, to address certain issues related to initialization of the DC programming-based method, we propose a more effective two-step approach which relies solely on convex programming.

\subsection{DC Programming Method}
We start by writing the task uploading energy
$\phi(\textbf{p},\textbf{b})$ introduced in (\ref{TotalEnergy}) as a DCF:
\begin{align}
\phi(\textbf{p},\textbf{b})&=\phi_1(\textbf{p},\textbf{b}) - \phi_2(\textbf{p},\textbf{b}) \label{UploadingEnergy}
\end{align}
where $\phi_1(\textbf{p},\textbf{b})=\sum_{j\in\mathcal{J}}(P_{j}+\frac{b_{j}}{2R_{j}})^2$ and $\phi_2(\textbf{p},\textbf{b})=\sum_{j\in\mathcal{J}}( P_{j}^2 + \frac{b^2_{j}}{4R^2_{j}})$. We also decompose the expected value of the total computation energy $\psi(\textbf{b},\tilde{\textbf{f}})$ as follows:
\begin{align}
\mathbb{E}_{\boldsymbol{\xi}}[\psi(\textbf{b},\tilde{\textbf{f}})]&=\mathbb{E}_{\boldsymbol{\xi}}\bigg[ \sum_{i\in\mathcal{I}}\kappa b_{i} c (1-\xi_i)^2f_{i}^2\bigg]\nonumber\\
&=\kappa c  \sum_{i\in\mathcal{I}}\eta_i \big( (b_{i} + f_{i}/2)^2 - (b^2_{i} + f^2_{i}/4) \big)\nonumber\\
&=\psi_1(\textbf{b},\textbf{f})-\psi_2(\textbf{b},\textbf{f})\label{ExpectationComptEnergy}
\end{align}
 where $\psi_1(\textbf{b},\textbf{f})= \kappa c  \sum_{i\in\mathcal{I}}\eta_i  (b_{i} + f_{i}/2)^2$, $\psi_2(\textbf{b},\textbf{f})= \kappa c   \sum_{i\in\mathcal{I}}\eta_i  (b^2_{i} + f^2_{i}/4)$ and $\eta_i=\mathbb{E}[(1-\xi_i)^2],\ i\in \mathcal{I}$. Finally, the objective function (\ref{Prob1}) expressed as a DCF:
 \begin{align}
\mathbb{E}_{\boldsymbol{\xi}}[E]&\equiv H(\textbf{x})=Y(\textbf{x})-Z(\textbf{x})\label{DC_objective}
\end{align}
where $Y(\textbf{x})=\phi_1(\textbf{p},\textbf{b}) +\psi_1(\textbf{b},\textbf{f})$ and $Z(\textbf{x})=\phi_2(\textbf{p},\textbf{b}) +\psi_2(\textbf{b},\textbf{f})$ are convex functions, and   $\textbf{x}=[\textbf{p}^\top  \textbf{b}^\top \textbf{f}^\top \textbf{t}^\top ]^\top$  contains all the search variables for convenience.

As shown in \cite{AlexanderS}, to apply a DC algorithm, each non-convex equality and inequality constraints can be incorporated into (\ref{DC_objective}) by using a penalty parameter once their DCFs are available. However, for our problem, we found that this approach yielded slow convergence. Whereas in \cite{AhmadiAA}, a DC algorithm is applied to a problem consisting of  only non-convex inequality constraints that are decomposed as DCFs. Hence, if we eliminate the equality constraint (\ref{C5}) by incorporating it into (\ref{DC_objective}) based on the penalty approach in \cite{AlexanderS}, we can develop a DC-based algorithm as in \cite{AhmadiAA} to solve a problem involving a penalized objective function (which is shown to be DCF \cite{AlexanderS}) with only inequality constraints (\ref{C3}) and (\ref{C4}).

To this end, we decompose the non-convex equality constraint (\ref{C5}) as:
\begin{align}
C^{\text{eq}}_j(\textbf{x})&=\dfrac{b_{j}}{R_{j}}-t^{\text{up}}_{j} =Y^{\text{eq}}_j(\textbf{x}) - Z^{\text{eq}}_j(\textbf{x}),\ \forall j\in\mathcal{J}\label{Det_C4_Decomposed_Final}
\end{align}
where $Y^{\text{eq}}_j(\textbf{x})=(b_{j}+ \frac{1}{2R_j})^2$ and $Z^{\text{eq}}_j(\textbf{x})=( b^2_{j} + \frac{1}{4R^2_j} + t^{\text{up}}_{j})$ are convex functions. Then we introduce the penalty term in the objective function, which can be also written as a DCF \cite{AlexanderS}:
\begin{align}
H_\lambda(\textbf{x})&=Y_\lambda(\textbf{x})-Z_\lambda(\textbf{x}) \label{PenalizedObj}
\end{align}
where
\begin{align}
Y_\lambda(\textbf{x})\hspace{-.07cm}&=\hspace{-.07cm}Y(\textbf{x})\hspace{-.07cm} +\hspace{-.07cm} 2\lambda\sum_{j\in\mathcal{J}}\max\big\{\hspace{-.07cm} Y^{\text{eq}}_j(\textbf{x});Z^{\text{eq}}_j(\textbf{x})\big\}\\
Z_\lambda(\textbf{x})&=Z(\textbf{x}) + \lambda\sum_{j\in\mathcal{J}} (Y^{\text{eq}}_j(\textbf{x})+Z^{\text{eq}}_j(\textbf{x})) 
\end{align}
and $\lambda\ge 0$ is the penalty parameter. 

In order to deal with the probabilistic inequality constraints (\ref{C3}) and (\ref{C4}), we adopt the chance-constraint programming approach \cite{Charnes}, and transform them into their deterministic equivalents. Specifically, constraint (\ref{C3}) can be given in terms of the cumulative distribution function (CDF) of $\xi_0$, $\textit{F}_{\xi_0}(\cdot)$, as follows:
\begin{align}
\mathbb{P}\bigg(\dfrac{b_{0}c}{(1-{\xi}_0)f_{0}}\hspace{-.06cm}\le\hspace{-.06cm} t^{\text{max}}\hspace{-.06cm}\bigg)\hspace{-.06cm}&=\hspace{-.06cm}\mathbb{P}\bigg(\hspace{-.05cm}{\xi}_0\hspace{-.06cm}\le \hspace{-.06cm}\underbrace{ \dfrac{f_{0}t^{\text{max}}-b_{0}c}{f_{0}t^{\text{max}}}}_{z}\bigg)\hspace{-.06cm}=\textit{F}_{\xi_0}(z)\ge  \gamma \nonumber
\end{align}
Then, assuming $\textit{F}_{\xi_0}(\cdot)$ is invertible, we can obtain the deterministic form of constraint (\ref{C3}) as follows:
\begin{align}
C_0(\textbf{x})=\dfrac{f_{0}t^{\text{max}}-b_{0}c}{f_{0}t^{\text{max}}}-F^{-1}_{\xi_0}(\gamma)\ge 0 \label{CDF_C3}
\end{align}
\noindent
where $\textit{F}^{-1}_{\xi_0}(\gamma)$ is the inverse CDF evaluated at $\gamma$. The new deterministic constraint (\ref{CDF_C3}) can now be written as a DCF in the following way:
\begin{align}
C_0(\textbf{x})=\text{ln}(\dfrac{b_{0}}{f_{0}}) - \text{ln}(q_0)=Y_0(\textbf{x}) - Z_0(\textbf{x})\le 0
\label{Det_C3}
\end{align}
\noindent
where  $q_0=t^{\text{max}}c^{-1}(1-F^{-1}_{\xi_0}(\gamma))$ is a non-negative constant, and $Y_0(\textbf{x})=-\text{ln}(f_0)$ and $Z_0(\textbf{x})=-\text{ln}(b_0) + \text{ln}(q_0)$ are convex functions.

Proceeding in a similar way, the deterministic form of constraint (\ref{C4}) is:
\begin{align}
C_j(\textbf{x}) =\dfrac{f_{j}(t^{\text{max}}-t^{\text{up}}_{j})-b_{j}c}{f_{j}(t^{\text{max}}-t^{\text{up}}_{j})}-F^{-1}_{\xi_j}(\gamma)\ge 0,\ \forall j\in\mathcal{J}
\label{Det_C4}
\end{align}
where $F^{-1}_{\xi_j}(\gamma)$ is the inverse CDF of $\xi_j$ evaluated at $\gamma$. In turn, (\ref{Det_C4}) can be decomposed as follows:
\begin{align}
\hspace{-0.2cm}C_j(\textbf{x})\hspace{-0.05cm}=\hspace{-0.05cm}\dfrac{t^{\text{up}}_{j}f_jq_j}{t^{\text{max}}}\hspace{-0.05cm}+\hspace{-0.05cm}b_{j}\hspace{-0.05cm}-\hspace{-0.05cm}f_jq_j\hspace{-0.05cm}=\hspace{-0.05cm}Y_j(\textbf{x})\hspace{-0.05cm}-\hspace{-0.05cm}Z_j(\textbf{x})\le 0,\hspace{-0.05cm}\ \forall j\hspace{-0.05cm}\in\hspace{-0.08cm}\mathcal{J}
\label{Decomposed_C4}
\end{align}
where  $q_j=t^{\text{max}}c^{-1}(1-F^{-1}_{\xi_j}(\gamma))\ \forall j\in\mathcal{J}$ is a non-negative constant, and $Y_j(\textbf{x})=\frac{q_j}{t^{\text{max}}}( 
t^{\text{up}}_{j} + f_j/2)^2 + b_{j}$ and $Z_j(\textbf{x})=\frac{q_j}{t^{\text{max}}}\big((t^{\text{up}}_{j})^2 + f^2_j/4\big) + f_jq_j$ are convex functions. 

Our first method is finally obtained by combining the DC programming approach in \cite{AhmadiAA} with the penalized DC approach in \cite{AlexanderS}; the resulting procedure is presented as Algorithm \ref{Iterative_PowerReducetion}. Following initialization, at the $k$th iteration of the algorithm, we first determine the convex approximations $H_\lambda^{(k)}(\textbf{x})$ and $C_i^{(k)}(\textbf{x})$ of $H_\lambda(\textbf{x})$ and  $C_i(\textbf{x})$ in step \ref{H_CONVEX} and \ref{C_CONVEX}, respectively, where $\nabla$ denotes the gradient operator. In step \ref{Solve_H}, we minimize $H_\lambda^{(k)}(\textbf{x})$ subject to the indicated constraints using standard convex optimization techniques until the sequence $\{H_\lambda^{(k)}(\textbf{x})\}$ converges with tolerance $\epsilon$ or the maximum iteration number $k^{\text{max}}$ is reached. The algorithm outputs the desired vector $\textbf{x}^{(k)}$ of the allocated resources.
 \begin{algorithm}[H]
	\caption{DC Algorithm Method}\label{Iterative_PowerReducetion}
	\begin{algorithmic}[1]
	\State $\textbf{input}$ Set $k=0$, initialize $\textbf{x}^{(0)}$
    \Repeat 
    	\State \hspace{-.4cm}$H_\lambda^{(k)}(\textbf{x})=Y_\lambda(\textbf{x})-Z_\lambda(\textbf{x}^{(k)})-\nabla Z_\lambda(\textbf{x}^{(k)})^\top(\textbf{x}-\textbf{x}^{(k)})  $\label{H_CONVEX}
    	\State \hspace{-.4cm}$C_i^{(k)}(\textbf{x})\hspace{-.1cm}=\hspace{-.1cm}Y_i(\textbf{x})\hspace{-.07cm}-\hspace{-.07cm}Z_i(\textbf{x}^{(k)})\hspace{-.1cm}-\hspace{-.1cm}\nabla Z_i(\textbf{x}^{(k)})^\top(\textbf{x}\hspace{-.1cm}-\hspace{-.1cm}\textbf{x}^{(k)}),\ i \in \mathcal{I} $\label{C_CONVEX}
    	 \State \hspace{-.4cm}solve $\textbf{x}^{(k+1)}=\argmin\limits_{\textbf{x}}H_\lambda^{(k)}(\textbf{x})$ \label{Solve_H}
    	 \State \hspace{2.2cm} $ \textrm{s.t.}\  C_i^{(k)}(\textbf{x})\le 0,\ i\in \mathcal{I}$
    	  \State \hspace{2.7cm} $ (\ref{C1}), (\ref{C2}), (\ref{C7})\ \text{and}\ (\ref{C8})$
	 \State \hspace{-.4cm} $k\leftarrow k+1$
	\Until{$|H_\lambda(\textbf{x}^{(k+1)})-H_\lambda( \textbf{x}^{(k)})|>\epsilon$ \textbf{or} $k\le k^{\text{max}}$}
    \State $\textbf{output}\ \textbf{x}^{(k)}$
	\end{algorithmic}
\end{algorithm}
\noindent

\subsection{Convex-Programming Method}
Although DC programming guarantees a local optimum by converging to a stationary point \cite{Vucic}, its performance depends heavily on the choice of the initial point $\textbf{x}^{(0)}$. To address this limitation, we propose a more  effective  two-step  approach relying solely on convex programming, which eliminates the dependence on user-defined initialization.

Consider an ideal scenario, in which there is no uncertainty in the allocated computation resources and the task uploading is instantaneous, i.e., $\xi_i=0, \forall i\in \mathcal{I}$ and $t^{\text{up}}_{j}=0, \forall j\in \mathcal{J}$. For this scenario let $\textbf{f}^{\star}$ and $\textbf{b}^{\star}$ be the optimal computation resources and the optimal task partitioning subject to constraint (\ref{C2}), which gives the minimum total energy consumption as $E^\star$. Note that based on (\ref{LocalCompTime}) or (\ref{TaskCompute}), we have  $f^{\star}_{i}t^{\text{co}}_i= b^{\star}_{i}c, \forall i\in\mathcal{I}$. It can be seen that at the optimal solution, the task completion time must match the given deadline, i.e., $t^{\text{co}^\star}_i=t^{\text{max}}$, since there cannot be any other computation resources, say $f_{i}^+\ \forall i\in \mathcal{I}$ with $f_{i}^+<f_{i}^\star$ that can reduce further the total energy consumption $E^\star$ without violating the time constraint or constraint (\ref{C2}). 

Based on the above, we can write the total computation energy in terms of only transmit power and task partitioning by replacing $t_0^{\text{co}}$ with $t^{\text{max}}$ and $t_{j}^{\text{co}}$ with $(t^{\text{max}}-t_{j}^{\text{up}})$ as follows:
\begin{align}
\psi(\textbf{p},\textbf{b})= \frac{\kappa (b_{0} c)^3 }{{(t^{\text{max}})}^2} + \sum_{j\in\mathcal{J}}  \frac{\kappa (b_{j} c)^3 }{\big(t^{\text{max}}-\frac{b_{j}}{R_{j}}\big)^2} \label{ComputationEnergy}
\end{align}
Hence, we decouple the allocation of computation resources and task partitioning in (\ref{ComputationEnergy}). More importantly, it can be shown that (\ref{ComputationEnergy}) is a convex function over the convex feasible set defined by constraints (\ref{C1}) and (\ref{C2}). Therefore, in the first step of our convex-programming method, we minimize the convex part $\psi(\textbf{p},\textbf{b})$ subject to constraints (\ref{C1}), (\ref{C2}) and a modified form of constraint (\ref{C5}) from problem $\mathscr{P}_1$:
\begin{subequations}
\begin{align}
\mathscr{P}_2:  \min_{\textbf{p}, \textbf{b}} \quad \psi(\textbf{p},\textbf{b}) \label{Prob2}\\
&\hspace{-2.5cm}\textrm{s.t.} \quad  0 \le \sum_{j\in\mathcal{J}}P_{j}\le P^\text{max} \label{P2_C1}\\
&\hspace{-1.73cm}   \sum_{i\in\mathcal{I}}{b_{i}} =b,\ b_i\ge0   \label{P2_C2}\\
&\hspace{-1.73cm} b_{j} - \alpha R_{j}t^{\text{max}}\le 0,  \forall j\in\mathcal{J} \label{P2_C3}%\\
\end{align}
\end{subequations}
Problem $\mathscr{P}_2$ can be easily solved by means of standard convex optimization methods. In constraint (\ref{P2_C3}), the scaling parameter $\alpha\in(0,1)$ is used to avoid the task uploading time exceeding the maximum time limit, i.e., $t^{\text{up}}_{j}>t^{\text{max}}$. In this way, constraint (\ref{P2_C3}) allows the computation time $t^{\text{co}}_{j}\ \forall j\in \mathcal{J}$ to be within the maximum time limit, and consequently, the solution of $\mathscr{P}_2$ lies in the feasible set of problem $\mathscr{P}_1$. We denote the solution of Problem $\mathscr{P}_2$ as $\textbf{b}^*$ and $\textbf{p}^*$.

In the second step, we minimize the expectation of the total computation energy (\ref{ExpectationComptEnergy}) with respect to computation resources subject to deterministic equivalents of constraints (\ref{C3}) and (\ref{C4}),  wherein the optimal values of $\textbf{b}^*$ and $\textbf{p}^*$ from Problem $\mathscr{P}_2$ are used in place of $\textbf{b}$ and $\textbf{p}$.
\\
\vspace{-.58cm}
\begin{subequations}
\begin{align}
\mathscr{P}_3:  \min_{\textbf{f}} \quad  
\mathbb{E}_{\boldsymbol{\xi}}[\psi(\textbf{b}^*,\tilde{\textbf{f}})&]= \sum_{i\in\mathcal{I}}\kappa  c \eta_i b^*_{i} f_{i}^2\label{Prob3}\\
&\hspace{-2.4cm}\textrm{s.t.} \quad 
\frac{b^*_{0}c}{t^{\text{max}}(1-F^{-1}_{\xi_0}(\gamma))}\le f_{0} \label{P3_C1} \\
&\hspace{-1.65cm} 
\frac{b^*_{j}c}{(t^{\text{max}}-t^{\text{up}^*}_{j})(1-F^{-1}_{\xi_j}(\gamma))}\le f_{j}, \forall j\hspace{-.8mm}\in\hspace{-.8mm}\mathcal{J}\label{P3_C2}
\end{align}
\end{subequations}
Note that $t^{\text{up}^*}_{j}=\frac{b^*_{j}}{R^*_{j}}\ \forall j\in\mathcal{J},$  where $R^*_{j}$ is the corresponding data rate for $P_j^*$; we then form the vector $\textbf{t}^*$ accordingly. It can be seen that the optimal solution $\textbf{f}^*$ of problem $\mathscr{P}_3$ can be directly calculated since it satisfies constraints (\ref{P3_C1}) and (\ref{P3_C2}) with equality.

After solving problem $\mathscr{P}_3$, the allocated computation resource at an  offloading device, say $j$, might exceed its computation capability, i.e., $f^*_{j}> f^{\text{max}}_{j}$. In this case, the solution of $\mathscr{P}_3$ is not in the feasible set of $\mathscr{P}_1$ as constraint (\ref{C7}) is violated. To address this issue, we reduce $f^*_{j}$ to $ f^{\text{max}}_{j}$ and we adjust the initially allocated task portion $b^*_{j}$ so that it can be computed without violating constraint (\ref{C7}). Specifically, we replace $b^*_{j}$ by 
\begin{align}
b^{+}_{j}=\dfrac{f^{\text{max}}_{j} R^*_{j}  t^{\text{max}} (1-F^{-1}_{\xi_j}(\gamma))}{R^*_{j}c+f^{\text{max}}_{j}(1-F^{-1}_{\xi_j}(\gamma))}\label{NewTaskSize}
\end{align}
which is  the maximum task portion size that can be computed by utilizing the full available computation resource $f^{\text{max}}_{j}$. The value of $b_{j}^+$
is obtained from constraint (\ref{P3_C2}) by replacing $f_{j}$ with $f^{\text{max}}_{j}$. Then the excess task portion, $b^{*}_{j}-b^{+}_{j}$, is assigned to the active device and/or the rest of the offloading devices. This is done by solving problem $\mathscr{P}_2$ and $\mathscr{P}_3$  after we remove the $j$th device from the set of available offloading destinations, i.e., we replace  $\mathcal{J}$ with $\mathcal{J}- \{j\}$. The process is repeated until constraint (\ref{C7}) is no longer violated by the remaining offloading devices.
If the set $\mathcal{J}$ becomes empty, then the leftover portion of the task size is computed at the active device, where we assume that $\frac{b^*_{0}c}{t^{\text{max}}(1-F^{-1}_{\xi_j}(\gamma))}= f^*_{0}< f^{\text{max}}_0$ based on constraint (\ref{P3_C1}). Finally, we present the overall progress of our second method in Algorithm \ref{ConvexProgramming}.

 \begin{algorithm}[h]
	\caption{Convex-Programming Method}\label{ConvexProgramming}
	\begin{algorithmic}[1]
	\State Solve $\mathscr{P}_2$ and $\mathscr{P}_3$ to obtain $\textbf{p}^*,\ \textbf{b}^*,\ \textbf{f}^* \text{and}\  \textbf{t}^*$
    \For{${j\in \mathcal{J}} $}\label{ForLoop}
    	\If{$f^*_{j}>f^{\text{max}}_j$}
        \State  \hspace{-0.5cm}	 $f^*_{j}\leftarrow f^{\text{max}}_j$
    	\State \hspace{-.5cm}	 Calculate the new task size $b^{+}_{j}$ using (\ref{NewTaskSize})
    	\State \hspace{-.5cm}	 $b^*_{j}\leftarrow b^{+}_{j}$
        \State \hspace{-.5cm}	 $b\leftarrow b-b^+_{j}$
        \State \hspace{-.5cm}	 $P^{\text{max}}\leftarrow P^{\text{max}}-P^{*}_{j
        }$   
        \State \hspace{-.5cm}	 Disregard the $j$th device: $\mathcal{J}\leftarrow \mathcal{J}-\{j\} $ 
        \State  \hspace{-.5cm}	 Update $\textbf{p}^*,\ \hspace{-.1cm}\textbf{b}^*,\ \hspace{-.1cm}\textbf{f}^* \text{and}\  \textbf{t}^*$ \hspace{-.15cm}  by \hspace{-.15cm}  solving \hspace{-.15cm} \hspace{-.1cm}  $\mathscr{P}_2$\hspace{-.05cm} and \hspace{-.15cm} $\mathscr{P}_3$
        \State \hspace{-.5cm}	 Go to line \ref{ForLoop} 
	    \EndIf\State \textbf{end}
	\EndFor \State \textbf{end}
    \State $\textbf{Output}\ \textbf{p}^*,\ \textbf{b}^*,\ \textbf{f}^* \text{and}\  \textbf{t}^*$
	\end{algorithmic}
\end{algorithm}
\noindent 

\section{Simulation Results}\label{Sec:SimResults}
In this section, we compare the energy efficiency and run-time of the proposed methods through Monte Carlo simulations. In each simulation run, we uniformly place the offloading devices on a disk with a radius set to $15$m centered at the active device. Furthermore, we consider independent Rayleigh fading channels and distance-dependent path loss model, $\text{PL}\hspace{-0.05cm}=\hspace{-0.05cm}148 + 40\mbox{log}_{10}(d)$ in dB, where $d$ is the distance in km \cite{YDAI}. As a benchmark  we also include the energy consumption 
when $\mathcal{J}\hspace{-0.05cm}=\hspace{-0.05cm}\emptyset$, i.e., the  task is completed locally. For the CPU throttling we assume that $\xi_i,\ i\in \mathcal{I},$ are uniform $\mathcal{U}(0,0.1)$, i.e., the actual computation resources may be below the allocated ones by up to $10\%$. We select the task size $b$ from a uniform distribution $\mathcal{U}(2\hspace{-0.06cm} \times\hspace{-0.06cm} 10^4\hspace{-0.05cm},\  \hspace{-0.1cm} 4\hspace{-0.06cm} \times\hspace{-0.06cm} 10^5 )$, and we set $f^{\text{max}}_0$ to a large value such that the assumption in the previous section holds. The rest of the system parameters are given in Table\hspace{-0.05cm} \ref{SysPar}.
\begin{table}[h]
	\vspace{0.17cm}
	\caption{System parameters} \label{SysPar}
	\resizebox{\columnwidth}{!}{%
		\begin{tabular}{ |l|c|c| }
			\hline
			\textbf{Parameter Description} & \textbf{Symbol} &\textbf{ Value} \\ \hline
			Number of offloading devices & $J$ & $ \{1, 2, 3\}	$ \\ \hline 
			CPU cycles to process 1-bit data & $c$ & $ 1500\ \text{cycles}/\text{bit}	$ \\ \hline	
	    	Effective  capacitance  constant & $ \kappa$ & $  10^{-24}\ $  W$\text{s}^3 $  \\ \hline
		    Max. iteration for DC prog. & $   k^{\text{max}}$ & $ 10^3	$ \\ \hline  
			Max.  time  constraint & $t^{\text{max}}$ & $ [.4, 1]\text{s}	$ \\ \hline  
			Max. transmit power & $	P^\text{max}$ & $  200\text{mW}	$ \\ \hline
			Limiting term for task uploading time & $ \alpha $ & $ .85$ \\ \hline
		    Reliability level  & $ \gamma $ & $ .95$ \\ \hline
		    Convergence tolerance for DC prog.  & $ \epsilon $ & $10^{-2} $ \\ \hline
		    Penalty parameter for DC prog.  & $ \lambda $ & $12$ \\ \hline
		    Max. radius of a D2D link  & - & 20m \\ \hline
		    Noise level & $N_0$ & $ -114	$dBm \\ \hline
		    Channel bandwidth & $W$ & $ 10	$MHz \\ \hline
		\end{tabular}	\label{SystemParameters}
	}
\end{table}
\begin{figure}[h]
	\centering
	\includegraphics[width=95mm]{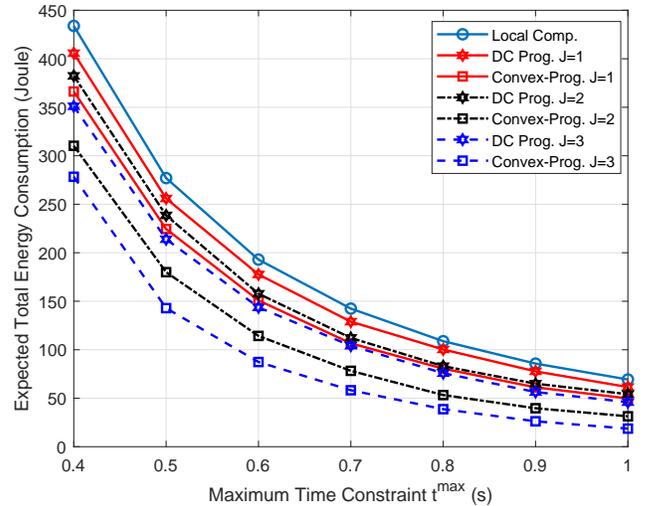}  
	\small
	\caption{Expected total energy consumption versus $t^{\text{max}}$ for different numbers of offloading devices ($\gamma=.95$).}
	\normalsize
	\label{TotalEnergyvsMaxTime}
\end{figure}

In Fig. \ref{TotalEnergyvsMaxTime}, we investigate the effect of the maximum time limit to complete the task. We assume that the devices have the same $t^{\text{max}}$ and to simulate different computation capabilities of the offloading device we select $f^{\text{max}}_j$ from a uniform distribution $\mathcal{U}(3\times10^7, 1\times10^8)$ for each simulation run. As seen from Fig. \ref{TotalEnergyvsMaxTime}, regardless of the time limit, both of our methods significantly reduce the total energy consumption compared to local task computation. However, the performance of the convex-programming method outperforms DC programming in terms of energy efficiency. Specifically,  $t^{\text{max}}=0.4$s, the total energy consumption with our convex-programming method requires almost 30\% less energy to compute the same task with respect to computing it only at the local device. Note that by increasing the maximum time limit, we can reduce the required computation resources, which naturally lowers the energy consumption. However, this negatively impacts the quality of service of the given task in terms of latency.

\begin{figure}[h]
	\centering
	\includegraphics[width=95mm]{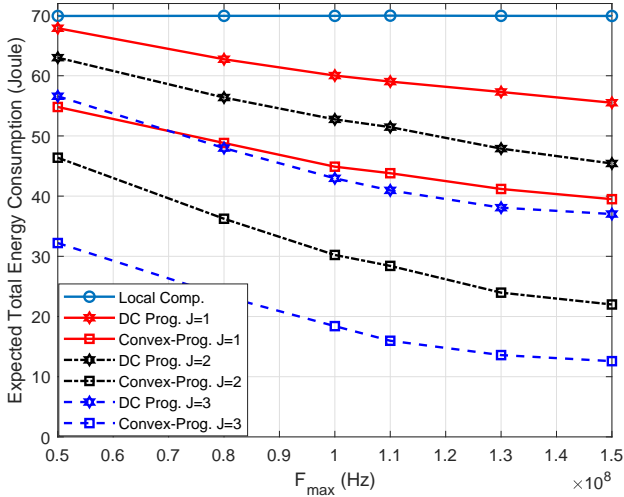}  
	\small
	\caption{Expected total energy consumption versus $\text{F}_{\text{max}}$ for different numbers of offloading devices ($t^{\text{max}}=1$,  $\gamma=.95$).}
	\normalsize
	\label{TotalEnergyVSMax_f_ES}
\end{figure}

In Fig. \ref{TotalEnergyVSMax_f_ES}, we consider the effect of maximum computation resources at the offloading devices on the total energy consumption. Specifically, we select $f^{\text{max}}_j$ from a uniform distribution $\mathcal{U}(\text{F}_{\text{min}}, \text{F}_{\text{max}})$, where $\text{F}_{\text{min}}$ is set to $3\times10^7$ Hz while $\text{F}_{\text{max}}$ is ranging from $5\times10^7-1.5\times10^8$ Hz. Even though increasing the number of offloading devices drastically reduces the energy consumption, the amount of available computation resources at the offloading devices limits the energy efficiency during task offloading. Therefore, reducing the total energy consumption not only depends on the number of nearby devices but is also highly affected by the amount of available computation resources that can be allocated by the offloading devices.

Finally, in Table \ref{RunTimeComparison} we compare the average run-time of proposed methods implemented in \textsc{Matlab} on an Intel i7-3770 computer with 16GB RAM. The proposed convex-programming based method not only achieves a better performance compared to our DC programming approach in terms of energy efficiency but also its run-time is significantly shorter. 
Specifically, with the increased number of offloading devices, DC programming takes at least ten times longer to converge within the selected tolerance value $\epsilon$. In addition, we observe that the required number of iterations for DC programming to converge is more than three times compared to our second method that is iteratively running Algorithm 2. 
\begin{table}[h!]
	\caption{Average run-time comparison ($t^{\text{max}}=.4$)} \label{RunTimeComparison}
	\resizebox{\columnwidth}{!}{%
		\begin{tabular}{ |l|c|c|c| }
			\hline
        	\multicolumn{2}{|c|}{\textbf{Simulation setup} }  & \multicolumn{1}{c|}{\textbf{DC Prog. Method} }& \multicolumn{1}{c|}{\textbf{Convex-Prog. Method}}  \\ \hline
		    $\text{F}_{\text{max}}=4\times10^7 $ & $J=1$ & 3.98 \text{s}  & 0.40 \text{s} \\ \hline 
		    $\text{F}_{\text{max}}=1\times10^8 $ & $J=1$ & 3.64 \text{s}  & 0.40  \text{s}  \\ \hline
		    $\text{F}_{\text{max}}=4\times10^7 $ & $J=2$ & 8.26 \text{s}  & 0.41  \text{s}  \\ \hline
		    $\text{F}_{\text{max}}=1\times10^8 $ & $J=2$ & 7.53 \text{s}  & 0.40  \text{s}  \\ \hline
		    $\text{F}_{\text{max}}=4\times10^7 $ & $J=3$ & 11.88 \text{s}  & 0.46 \text{s}  \\ \hline
		    $\text{F}_{\text{max}}=1\times10^8 $ & $J=3$ & 11.60 \text{s}  & 0.42 \text{s}  \\ \hline
		\end{tabular}	
	}
\end{table}

\section{Conclusion}\label{Sec:Conclusion}
In this paper, we proposed two sub-optimal methods for a D2D-aided fog computing scenario under probabilistic time constraints. The first method relies on DC programming, however, its performance is very sensitive to the choice of the initial point. Hence, we propose a novel alternative solution based on convex programming, which eliminates the dependence on user-defined initialization. Nevertheless, due to the uncertainties on the allocated computation resources, we incorporate chance-constraint programming into both methods. While both proposed sub-optimal task offloading methods significantly reduce the total energy consumption compared to computing the task locally, the second method outperforms DC programming in terms of energy efficiency and run-time.

\end{document}